# Natural Disasters, Entrepreneurship Activity, and the Moderating Role of Country Governance

| **Christopher J. Boudreaux** | **Anand Jha** | **Monica Escaleras** |

Florida Atlantic University
College of Business

777 Glades Road, Boca
Raton, FL 33431
cboudreaux@fau.edu

Wayne State University
Mike Ilitch School of Business

2771 Woodward Ave,
Detroit, MI 48201
anand.jha@wayne.edu.

Florida Atlantic University
College of Business

777 Glades Road, Boca
Raton, FL 33431
mescaleras@fau.edu

## ABSTRACT

The purpose of this paper is to investigate if a country's quality of governance moderates the effect of natural disasters on start-up activity within that country. We test our hypotheses using a panel of 95 countries from 2006 to 2016. Our findings suggest that natural disasters discourage start-up activity in countries that have low-quality governance but encourage start-up activity in countries that have high-quality governance. Moreover, our estimates reveal that natural disasters' effects on start-up activity persist for the short term (1–3 years) but not the long term. Our findings provide new insights into how natural disasters affect entrepreneurship activity and highlight the importance of country governance during these events.

## PLAIN ENGLISH SUMMARY

Natural disasters encourage more start-up activity, but only in countries that have high-quality governance. In countries with low-quality governance, natural disasters discourage start-up activity. Moreover, our estimates reveal that natural disasters' effects on start-up activity persist for the short term (1–3 years) but not the long term. Our findings provide new insights into how natural disasters affect entrepreneurship activity and highlight the importance of country governance during these events. These findings have important implications for both researchers and policymakers. Researchers should note the effect of natural disasters on entrepreneurship activity is nuanced and contingent upon country governance. Our study is also useful to policymakers who want to limit the adverse impact of natural disasters on start-up activity. Policymakers recognize these new firms are the backbone of new job creation, and they are more likely to invest in potential innovative breakthroughs, leading to more employment in the future.

# 1. INTRODUCTION

In the 1970s, there were fewer than 100 natural disasters in a year. In 2010, this number approached 400. Although the number of deaths due to natural disasters is decreasing, losses due to these disasters are not.[1] According to a report by the World Bank dedicated to understanding the financial and economic impacts of natural disasters, "Between the 1950s and the 1990s, the reported global cost of natural disasters increased 15-fold. Major natural catastrophes in the 1990s caused economic losses estimated at an average $66 billion per year (in 2002 prices)" (Benson and Clay 2004). In 2018, natural disasters cost the world $160 billion.[2]

Despite the increasing costs that natural disasters impose on an economy, academic research has seldom investigated how to mitigate the negative impact of a natural disaster. However, one notable exception has examined whether the country's governance quality mitigates the adverse impact of disasters on multi-national corporation (MNC)'s subsidiary investment (Oh and Oetzel 2011). The authors examined the impact of different types of disasters—including natural disasters—on the MNC's subsidiary investment to investigate whether the quality of governance mitigates the adverse effect of disasters on the MNC's investment.[3] They found that when the host country has better regulatory quality, natural disasters appear to increase the number of subsidiaries, suggesting that high-quality regulations help MNCs exploit business opportunities arising after a natural disaster. Other aspects of governance (government effectiveness, the rule of law, voice and accountability, political stability and control of corruption) did not matter. One explanation is that MNCs are in a better position to deal with a natural disaster, they are likely to be insured, and they often have the option of channeling resources from other parts of the world to exploit opportunities that natural disasters create.

---

[1] People are increasingly living and buying property in natural disaster-prone areas as they feel safer due to better technology to predict the timings of these disasters and the availability of insurance (Sadowski and Sutter 2005).
[2] https://weather.com/science/environment/news/2019-01-09-disasters-cost-damage-climate-change
[3] They measure quality of governance using the World Governance Indicators (government effectiveness, regulatory quality, the rule of law, voice and accountability, political stability and control of corruption). We describe these measures in more detail in the data section.

In contrast with MNCs, new entrepreneurs often have limited resources and are thus ill equipped to deal with disasters. Natural disasters not only disrupt networks and destroy infrastructure, making day-to-day operations costly (Boehm, Flaaen, & Pandalai-Nayar, 2018), but they also present new opportunities (Cuaresma, Hlouskova, & Obersteiner, 2008). And we know from prior literature that the quality of institutions incentivizes entrepreneurs to start new businesses (Nikolaev et al. 2018; Baumol 1990; Sobel 2008; Estrin et al. 2013). Whether entrepreneurs can exploit the opportunities and limit the costs associated with natural disasters may depend on the quality of governance in the country. High-quality governance might lessen a natural disaster's adverse impact and grease its positive impact on starting a new business. For example, if a country's residents are free to voice public concerns, entrepreneurs will anticipate a quicker recovery from the damage done to infrastructure. Moreover, entrepreneurs might anticipate more relief material and better access to credit with more effective governance. If a country's governance facilitates disaster recovery and greases entrepreneurs' desires to exploit opportunities, we expect the number of start-ups to increase following a natural disaster in a country with high-quality governance. However, we expect no such effect in countries that lack good governance.

Motivated by this line of thought, and with the goal of understanding how a country's quality of institutions might affect start-up activity following a natural disaster, we begin by investigating the association between natural disasters and the rate of start-up activity. Next, we investigate whether the quality of governance affects the rate of start-up activity. We then focus on the central question of this study: does the quality of governance in a country moderate the relationship between natural disasters and start-up activity? We hypothesize the adverse effect of natural disasters on the rate of start-up activity is more severe when countries have low-quality governance, but that the effect is less severe or even positive when countries have high-quality governance. We develop hypotheses using three strands of literature regarding entrepreneurship, natural disasters, and the role of country-level governance in fostering business activity. To test these hypotheses, we collected data from the World Bank's Entrepreneurship Survey and Database, which provides data on the annual rate of new start-ups in a country. Next, using data form the Emergency Events Database we constructed a measure of natural disaster intensity by adding the total number of people affected, injured, and made homeless due to a natural disaster, and then merged this variable with the World Bank data. Finally, we merged these variables with measures of governance from the World Governance Indicators (government effectiveness, regulatory quality, rule of law, voice and accountability, political stability, and control of corruption). Our sample is comprised of 95 countries from 2006 to 2016 and includes countries with high-quality institutions such as Norway, Finland, and Switzerland as well as countries with low-quality institutions such as Saudi Arabia, Belarus, Laos, and Tajikistan.

Our results reveal a negative but typically insignificant relationship between natural disaster intensity and start-up activity. We found that only two country governance characteristics—voice and accountability and regulatory quality—have a positive association with start-up activity. More importantly, we tested and found support for the central hypothesis of our study: the quality of governance in a country moderates the effect of natural disasters on start-up activity. Specifically, high-quality governance *positively* moderates the association between natural disasters and entrepreneurship start-up activity. In other words, when countries have low-quality governance, we cannot expect entrepreneurs to establish a new business promptly following a disaster. However, when a country's governance is of higher quality, entrepreneurs can take advantage of the business opportunities a disaster might create.

These findings are important for several reasons. First, our findings are useful for entrepreneurs who are eager to exploit the opportunities that natural disasters create. Our study offers an idea of what to expect following a natural disaster given the quality of governance in the country. Second, our study is useful to policymakers who want to limit the adverse impact of natural disasters on start-up activity. Policymakers recognize that these new firms are the backbone of new job creation,[4] and they are more likely to invest in potential innovative breakthroughs, leading to more employment in the future. Finally, more business start-ups after a natural disaster might help the economy to recover more quickly. We discuss these findings and their policy implications in more detail later in the paper.

## 2. THEORY AND HYPOTHESES
### 2.1. Natural disasters and entrepreneurship activity

It is unclear *a priori* exactly how natural disasters affect start-up activity. On the one hand, crises create opportunities. As scholars have pointed out, "The Chinese symbol for crisis combines two simpler symbols—the symbol for danger and the one for opportunity. Crises are times of danger, but they are also times of opportunity" (Starbuck et al. 1978). Winston Churchill reportedly said, "Never let a good crisis go to waste."[5] It is therefore not surprising that entrepreneurs often view a crisis as an opportunity for new business ventures (Brück et al. 2011). Research on the topic highlights how loan demand spikes after a natural disaster as people attempt to recover their losses. This lending activity increases consumption and investment, increasing general business activity in the process. This can create room for new types of businesses; hence, the number of start-ups might increase (Monllor and Murphy 2017).

---

[4] Between 1977 and 2009 roughly two to three millions new jobs (almost all of the new jobs) were created by startups in the U.S. https://www.forbes.com/sites/petercohan/2011/06/27/why-start-ups-matter/#59238e143620
[5] https://realbusiness.co.uk/as-said-by-winston-churchill-never-waste-a-good-crisis/

On the other hand, uncertainty often accompanies natural disasters. Given this uncertainty, entrepreneurs find it more difficult to navigate *disequilibria* (Schultz 1975) which increases the risk of failing and might discourage new start-ups. For example, a natural disaster can result in the breakdown of supply chain networks (Carvalho 2014), affect the safety of workers, make it difficult to return to normal operation (Chamlee-Wright and Storr 2009; Grube and Storr 2018), and therefore decrease productivity (Boehm et al. 2019). Although new opportunities might be available, the cost of taking advantage of those opportunities might increase. Research shows that it can be difficult for entrepreneurs to receive a loan. Berg and Schrader (2012), for example, have found that although the demand for credit increases after a disaster, just when many households want to borrow to cope with the disaster, their access is restricted. The risk preference of those affected by the disaster might also change. Cassar et al. (2017), who conducted experiments with 334 Thai subjects in the province worst affected by the 2004 tsunami, have found that natural disasters can make people more risk averse. This has important ramifications for entrepreneurship, which is inherently risky (Knight 1921).

Despite these theoretical underpinnings, the empirical literature on this topic is sparse. The only study we are aware of that has tested the effect of natural disasters on entrepreneurship activity is Boudreaux et al. (2019a; 2021) which found that natural disaster events decrease entrepreneurship activity in the year immediately following a natural disaster. There are a few related studies. For example, Escaleras and Register (2011) have examined the impact of natural disasters on foreign direct investment. They have found that natural disasters decrease foreign direct investment. Another related study has found that natural disasters have no impact on MNCs' numbers of subsidiaries (Oh and Oetzel 2011). Natural hazards can devastate physical capital, labor stocks, and the social infrastructure necessary for commerce (e.g., transportation and communication networks).

We argue that the uncertainty effect will likely dominate the increased opportunity effect, and therefore we make the following hypothesis:

*H1: Natural disasters decrease start-up activity.*

### 2.2. Country governance and entrepreneurship activity

Good governance at the country level affects business activity, and economists are increasingly interested in understanding its role in the economy. In his presidential address to the American Economic Association in 2009, Avinash Dixit mentioned that "EconLit shows only 5 mentions of the word governance in the 1970s; by the end of 2008, it was mentioned 33,177 times" (Dixit 2009).

Kaufmann et al. (1999) have postulated six different aspects of good governance: government effectiveness, regulatory quality, rule of law, voice and accountability, political stability, and control of corruption. An effective government means that the public and civil service are largely independent of political pressures in implementing policy. High-quality regulation means that the rules formulated by the state promote private sector development. The rule of law implies that the public has confidence that everyone will be treated equally under the law. In countries with a higher-quality rule of law, citizens perceive that by and large, contracts are enforced, property rights are valued, and courts provide justice. The level of criminal activity is also lower. For a country to have voice and accountability means that the citizens can express themselves freely and have access to free media. Political stability means that there is less politically motivated violence and terrorism within a country. Finally, lower corruption implies that politicians from well-governed countries are less likely to use their power to expropriate rents from business, bribery is less common, and the elite are less likely to capture the state. Though these different aspects of good governance are distinctly different concepts, they are interrelated. For example, an effective government is also less corrupt and has more inbuilt checks and balances, which nurtures the rule of law.

Dixit (2009) has explained why country-level governance is necessary for a properly functioning market economy. He has posited good governance is associated with the security of property, enforcement of a contract, and facilitation of collective action. When property rights protection is weak, individuals fear others will take the fruits of their endeavors. They also spend more time guarding their property. In countries with lax contract enforcement, there is a higher probability one of the participants in the business endeavor will act opportunistically. Similarly, a lack of infrastructure to facilitate collective action (e.g., quick justice) such as punishment for deviant behavior (e.g., embezzlement, extortion) deters the proper functioning of the market since good business requires cooperation from many players. Unsurprisingly, researchers in international business have theorized poor quality of institutions is associated with market failures (Peng et al. 2008) and influences an MNC's business strategy (Pinkse and Kolk 2012).

Research has specifically examined the connection between entrepreneurship and country-level governance. Webb et al. (2020) theorize about the effect of formal and informal institutional voids on entrepreneurship and suggest that a lack of quality formal institutions can discourage legal tax-paying entrepreneurship. Specifically, they posited that "in societies characterized by more severe formal institutional voids, entrepreneurs can incur relatively high costs in gaining formal status but receive minimal benefits in return; and given minimal constraints imposed on pursuing more informal activities, entrepreneurs can create value for themselves with relatively less risk" (page 512). Mickiewicz and Olarewaju (2020) conducted a case study of a farm run by migrants in Nigeria and noticed that cost of doing business is higher when there is a lack of quality institutions. Anokhin and Wincent (2012) have used an indirect proxy measure of country governance (i.e., the stage of development) and found that it positively moderates the effect of start-up rates and innovation, suggesting it is easier to reap the benefits of entrepreneurship when country governance is of high quality. Aidis et al. (2008) have blamed weak institutions for stunting start-up activity in Russia. Using data from over 70 countries, Nikolaev et al. (2018) have demonstrated that institutional variables—particularly those related to economic freedom—are strongly correlated with entrepreneurial activity.[6]

Based on the aforementioned studies (e.g., Nikolaev et al. 2018), we believe that low-quality country-level governance measures lead to greater uncertainty and ultimately higher costs of capital. Holding other variables in the net present value (NPV) calculation constant, a higher cost of capital reduces the NPV, and a project that would have otherwise been undertaken is not. Empirical research supports this conjecture (Globerman and Shapiro 2003; Oh and Oetzel 2011).[7]

Start-ups are particularly sensitive to uncertainty because they have fewer resources, they are opaquer, and they have fewer tools to navigate the ineffectiveness of the government and

---

[6] There are many other studies that examine the association between the institutional context and entrepreneurship activity and conclude that quality institutions can make it easier to foster new business. (e.g., Bjørnskov and Foss 2016; Boudreaux 2014; Boudreaux, et al. 2019b; Chowdhury et al. 2019; Estrin et al. 2013; Hwang and Powell 2005; Stephan et al. 2015; Urbano et al. 2019).

[7] Globerman and Shapiro (2003) find that countries that fail to meet the minimum threshold of good governance are unlikely to receive any FDI from U.S. multinational corporations. Further, among the firms that do receive FDI, they find the ones with better governance receive more FDI. Oh and Oetzel (2011) Oh and Oetzel (2011) find that all six measures of country-governance are associated with more foreign subsidiaries. Unsurprisingly, countries with good governance are more productive and grow faster (Keefer and Knack 1997; Hall and Jones 1999).

cumbersome regulations and satisfy the demands of powerful public officials. They also have more difficulty raising capital compared to MNCs. For these reasons, we propose the following hypothesis:

*H2: High-quality governance increases start-up activity.*

## 2.3. Quality of governance moderates the relationship between natural disasters and entrepreneurship activity

Entrepreneurs expect a country with high-quality governance to quickly recover from a natural disaster.[8] When a country has an effective, stable, and less corrupt government, it is easier to coordinate and organize for the common good, and the recovery process can proceed much more smoothly. Berke et al. (1993) have reviewed the key findings in the literature regarding disaster recovery. They have summarized that the key to better recovery is strong structural and functional relations among various social units and the capacity to "diffuse, adapt, implement plans, and policy innovation" (page 107). In countries that have an ineffective, corrupt government, political interference often delays policy implementation. Moreover, the communication between various units is often problematic because these countries lack the infrastructure for collective action (Dixit 2009). Boudreaux et al. (2021) find the quality of institutions moderates the effect of foreign aid on starting a new business after a disaster.

Low-quality governance compromises the proper deliberation, planning, and implementation of disaster recovery. Johnson and Olshansky (2017) have drawn on decades of research on disaster recovery and compared disaster recovery efforts in China, Japan, New Zealand, India, Indonesia, and the U.S. They have also compared responses to earthquakes in

---

[8] The market process alone cannot itself lead to a faster recovery process unless the government plays an active role (Chamlee-Wright 2010; Sutter 2011).

China and Japan. In China, a 7.9 magnitude earthquake killed 70,000 people and displaced 1.5 million in 2008. China's government tried to reconstruct everything in three years in an effort that was largely led by the central government without proper deliberation. The quality of recovery suffered as a result. High-profile developments fell apart before completion. Rural residents lost their livelihoods as the government relocated them to cities. In contrast, after Japan's 1995 earthquake, local government largely led the recovery effort using funding from national leaders. The process involved many vigorous deliberations, and the recovery efforts were not made hastily. The infrastructure the local governments developed is earthquake-proof, with considerably fewer costs to residents. It is reasonable to argue that the free press, the rule of law, and proper enforcement of implicit and explicit contracts between the Japanese government and its people lead to proper deliberation and planning.

Thus, in countries with high-quality governance, entrepreneurs are likely to expect less chaotic recovery that involves proper planning and implementation. Hence, an entrepreneur's risk assessment of a future business opportunity will be lower. Furthermore, in countries with high-quality governance, private endeavors to exploit the opportunities after a disaster are likely to proceed much more quickly without political interference. These countries have regulations that encourage private sector development. Therefore, entrepreneurs are less likely to fear political extortion when establishing a new business.

Together, the lower assessment of risk and lower implicit and explicit costs of starting a business should encourage entrepreneurs to establish new start-ups as demand increases following a disaster. Based on this reasoning, we posit the following hypothesis:

*H3: High-quality country governance positively moderates the effect of natural disasters on the start-up activity rate.*

# 3. DATA AND METHODS
## 3.1. Data sources and sample

To test our hypotheses, we gathered data from a variety of sources. We began our sample by collecting data on 1,831 new business start-up rates from the World Bank's Entrepreneurship Survey and Database from 2006 to 2016. We then merged this sample with country governance data from the World Governance Indicators (WGI). The WGI include 2,169 observations over the sample period, but only 1,280 observations remained after the merge with the World Bank data. Next, we merged this sample with natural disaster data from the Emergency Events Database (EM-DAT). The EM-DAT includes 1,881 observations over the sample period, but only 1,162 observations remained after the merge. Finally, we merged our sample with our set of control variables gathered from the World Bank and the World Bank's Doing Business Survey. After the merge, 701 observations remained. This final sample includes data from 2006 to 2016 from a panel of 95 countries. We report the definitions and data sources for each variable in Table 1 and report the summary statistics of the key variables by country in Table 2.

It is important to note that we follow recent calls in the entrepreneurship and management literature urging scholars to focus less on statistical significance and more on effect sizes (e.g., Anderson et al. 2019; Meyer et al. 2017). Specifically, we report exact p-values and do not report asterisks in regression tables indicating statistical significance. However, we use a $p < 0.10$ threshold for hypothesis testing. Lastly, we report our effect sizes graphically using 95% confidence intervals.

---------------------------------
Insert Table 1 About Here
---------------------------------

---------------------------------
Insert Table 2 About Here
---------------------------------

**3.2. Dependent variable**

Our dependent variable, the new business start-up rate, is the number of newly registered limited-liability firms as a percentage of a country's working-age population (ages 15–64) normalized by 1,000. This variable comes from the World Bank's Entrepreneurship Survey and Database. We used a histogram to examine this variable's distribution and found that it was power-law distributed[9], which is consistent with recent evidence from entrepreneurship studies (Crawford et al. 2014; Crawford et al. 2015). As a result, we used the natural logarithm to transform this measure, which resulted in a normal distribution (see Figure 1).

---------------------------------
Insert Figure 1 About Here
---------------------------------

**3.3. Explanatory variables**

We are particularly interested in two explanatory variables: natural disasters and country governance. The data on natural disasters came from the EM-DAT maintained by the Centre for Research on the Epidemiology of Disasters (CRED). A disaster is included in the database if it satisfies at least one or more of the following criteria: (1) 10 or more persons killed; (2) 100 or more persons affected, injured, or left homeless; (3) an appeal for international assistance; or (4) an official declaration of a state of emergency. Our analysis includes the following natural disasters: earthquakes, floods, slides, volcanic eruptions, and windstorms. Disaster data are available from 1900 to present, but we only used the data that overlapped with the data from our sample period (2006–2016). Most studies, including ours, that attempt to quantify the intensity of the disasters use a measure based on the number of people affected, injured, or killed from EM-DAT (for example, see Loayza et al. 2012; Klomp and Valckx 2014; Noy 2009; Boudreaux et al. 2021). Therefore, we define the "intensity of natural disaster" as the sum of the total number of people affected, injured, and homeless due to a natural disaster.[10] We transform this variable using the natural logarithm and add one to account for observations with zeros.[11]

---
[9] Power law distribution refers to a highly skewed distribution where a few outliers account for a disproportionate amount of the total distribution's output (Crawford et al., 2015).

We gathered data from the WGI to measure our variable: country governance. The WGI measure country governance using six different measures: voice and accountability, political stability, government effectiveness, regulatory quality, rule of law, and control of corruption. These six indicators are based on the opinions of country experts and are scaled from -2.5 to 2.5 with a standard normal distribution (mean = 0; SD = 1). For these indicators, higher scores reflect higher-quality country governance and lower scores reflect lower-quality country governance. To increase the robustness of our findings, we included each of the six measures in our models. Table 1 provides more details regarding each variable's definition.

### 3.4. Control variables

In addition to our explanatory variables, we also included several control variables to mitigate omitted variable bias concerns. Including these control variables adjusted for their effects on our explanatory variables, dependent variables, or both. We gathered these control variables from the World Bank's Doing Business Survey. First, we gathered data on economic and financial indicators (GDP per capita, GDP growth, financial credit, and trade) that influence entrepreneurship activity or country governance. Economic indicators like GDP per capita and GDP growth capture economic activity in each country. We used the natural logarithm to transform GDP per capita, and we measured GDP growth as the annual growth rate. Gross domestic product per capita is thus the log of real GDP per capita using purchasing power parity, and GDP growth is measured as the annual percentage change in GDP. Similarly, our financial indicator, financial credit, captures a country's economic activity since it measures the domestic credit provided by the financial sector. This variable is expressed as a percentage of GDP. Our final economic indicator, trade, captures the trade openness of an economy. Countries with larger amounts of trade with others, which is expressed as a percentage of GDP, are likely to be more developed.

---

[10] Affected are those requiring immediate assistance during a period of emergency. Injured are those suffering from physical injuries, trauma, or an illness requiring immediate medical assistance as a direct result of a disaster. Homeless are those people whose house is destroyed or heavily damaged and therefore need shelter after an event.

[11] Log (0) is undefined so it will be dropped. Therefore, we use the transformation, Log (0+1) = log(1) = 0.

We also adjusted our model to reflect the fact that different countries have varying entry regulations associated with starting a new business based on work by Djankov et al. (2002).These variables capture the ease (or difficulty) of starting a new business. We gathered these variables from the World Bank's Doing Business Survey. The cost of business start-up procedures measures the total cost required to complete these procedures as a percentage of gross national income (GNI) per capita. The time required to start a business is measured in the number of days. Finally, the number of start-up procedures measures the total number of procedures required to register a business. The idea behind these Doing Business indicators is that we should expect more business creation when it is less costly, requires less time, and requires fewer procedures. De Soto (2000) originally found that it took his team 278 eight-hour days to open a business in Peru. This variable ranges from half a day in New Zealand to 690 days in Suriname.

Finally, we included three demographic variables that might influence either entrepreneurship activity or country governance. The largest population was measured as the largest city population expressed as a percentage of the country's total population. Population density measures the people per square kilometer, which we defined as the population divided by the land area. Land area measures the area of land in square kilometers using the log transformation. These three demographic indicators might influence entrepreneurship activity, such as if entrepreneurship is more likely to be located in densely populated and urban areas.

---------------------------------
Insert Table 3 About Here
---------------------------------

Table 3 reports the descriptive statistics and correlation matrix for these variables. There is a negative and significant correlation between natural disasters and new business start-up activity and significant positive correlations between our six country governance indicators and new business start-up activity. This provides some preliminary evidence to support Hypothesis 1 and Hypothesis 2, but we require more sophisticated tests before reaching that conclusion. Additionally, it takes an average of 23 days to start a business with almost eight procedures, which costs roughly 28 percent of GNI. Clearly, starting a business is not a trivial task in many locations. There are also significant correlations between our six governance indicators. While many of our variables are modestly correlated, GDP per capita and the six governance indicators are often very strongly correlated. In additional robustness tests (Appendix, Table A1), we excluded GDP per capita from our model and found the results to be qualitatively similar.

**3.5. Model**

Our empirical model can be represented as follows:

$$STARTUP_{it} = \alpha + \beta_1 DISASTERS_{it-1} + \sum_{j=1}^{6} \gamma_j GOVERN_{it-1}$$

$$+ \sum_{j=1}^{6} \delta_j (DISASTERS_{it-1} \times GOVERN_{it-1}) + \sum_{k=1}^{10} \mu_k CONTROLS_{it}$$

$$+ \sum_{l=1}^{7} \eta_l REGIONS_l + \sum_{t=2006}^{2016} \lambda_t YEAR_t + \varepsilon_{it},$$

where STARTUP is the natural logarithm of the new business start-up rate (the number of new firms divided by the working age population) in a country, DISASTERS is the natural logarithm of natural disaster intensity, GOVERN is a vector of the six country governance indicators, DISASTERS × GOVERN is a vector of interactions for each of the six country governance indicators, CONTROLS is a vector of the 10 previously described control variables, REGIONS is a vector of region dummies, YEAR is a set of year dummies, and ε is an idiosyncratic error term. We included region and year dummies to adjust for region-specific and year-specific differences between countries and over time.[12] We estimated this model using linear regression models (i.e., ordinary least squares). These models assume homoscedasticity where the error term is independently and identically distributed. We adjusted for this assumption in two ways. First, our dependent variable is log-transformed, which greatly reduces heteroscedasticity concerns. Second, we used White's robust standard errors, which are consistent in the presence of heteroscedasticity (White 1980).

---

[12] Consistent with Oh and Oetzel (2011) we do not use country fixed effect as country-governance do not change much over a span of 10 years and therefore using country-fixed effect would throw away all the time invariant component of governance measure. It would be equivalent what some academic call "throwing the baby out of the water (Angrist and Pischke 2009)"

## 4. RESULTS

We present the results of our regression model in Table 4. Column (a) of Table 4 presents our baseline model, which only includes control variables. Financial and economic activity (GDP per capita, GDP growth, and financial credit) are positively associated with start-up rate. We also found that regulatory barriers to start-up activity (cost of business start-up procedures, time required to start a business, number of start-up procedures) are negatively associated with start-up rates. This result is consistent with studies that have found substantial entry barriers to entrepreneurship, especially in developing economies (De Soto 2000; Djankov et al. 2002). Finally, we found that larger and more densely populated countries have lower start-up rates.

---------------------------------
Insert Table 4 About Here
---------------------------------

### 4.1. Disasters and entrepreneurship activity (Hypothesis 1)

Column (b) of Table 4 augments our baseline model to include our measure of natural disasters. While there is a negative sign on the natural disasters' coefficient, it is not statistically significant ($\beta$ = -0.008; $p$ = .361). Moreover, a likelihood ratio test that compared the log-likelihoods of models (a) and (b) indicated that natural disasters have no significant effect on new business start-up rates. That said, although natural disasters are never statistically significant across Table 4's specifications, the coefficient of natural disaster is consistently negative. We therefore explore this relationship in more detail in Table 5 and in the discussion section.

### 4.2. Quality of governance and entrepreneurship activity (Hypothesis 2)

Columns 1 through 6 of Table 4 test our hypothesis that high-quality country governance is positively associated with entrepreneurship activity. In these columns, each of the six measures of country governance from the WGI estimate the effect of quality of governance on the rate of start-up activity. Four of the six indicators are positively associated with the rate of start-up activity, and the remaining two indicators are negatively associated with the rate of start-up activity. However, only two indicators—voice and accountability (β = 0.351; $p$ = .000) and regulatory quality (β = 0.269; $p$ = 0.008)—have both positive and statistically significant associations with the rate of start-up activity. Thus, while there is some evidence to support Hypothesis 2, it is not fully supported. We therefore explore this relationship in more detail in Table 5 and discuss the implications in the discussion section.

### 4.3. Moderating effects of quality of governance (Hypothesis 3)

In Table 5, we included an interaction term between natural disasters and the six governance indicators. These models test our hypothesis that a country's quality of governance moderates the relationship between natural disasters and entrepreneurship activity. We found overall support for Hypothesis 3; that is, the results suggest that natural disasters discourage entrepreneurship activity, but the quality of governance attenuates this adverse effect. For example, Column 1 of Table 5 reports that natural disasters are negatively (though not significantly) associated with the new business start-up rate (β = -0.008; $p$ = .339), but voice and accountability (β = 0.014; $p$ = .064) attenuates this effect. There is a very consistent relationship for each of the remaining governance indicators (political stability, government effectiveness, regulatory quality, rule of law, and control of corruption). These results indicate that in countries where the quality of governance is higher, natural disasters have a positive effect on new business start-up activity. In other words, the number of start-ups increases following a disaster when countries have good governance quality, but the number of start-ups decreases following a disaster when countries have lower governance quality. Thus, Hypothesis 3 is supported.

---------------------------------
Insert Table 5 About Here
---------------------------------

To ease the interpretation of this moderation, we plotted the marginal effects of these models in Figures 1 through 3. Our model reports elasticity estimates since we estimated a model in the form $\ln(y) = a + b * \ln(x)$. We thus interpreted the marginal effects as a one percent increase in the explanatory variable is associated with a percentage increase in the outcome variable. Figure 2 illustrates the moderating effects using the governance indicators of voice and accountability and political stability. The results suggest that both indicators significantly moderate the relationship between natural disasters and new business start-up activity. In countries with low-quality voice and accountability scores, natural disasters have a negative effect on new business start-up activity. However, as the quality of voice and accountability increases, the negative effect becomes smaller or even positive. We observed a very similar relationship using the political stability governance indicator. Natural disasters have a negative effect on new business start-up activity, but higher-quality political stability governance attenuates this effect. Specifically, our estimates suggest that a ten percent increase in natural disaster activity is associated with a 0.66 percent decrease in new business start-up activity when voice and accountability is low (-2.5). However, a ten percent increase in natural disaster activity is associated with a 0.38 percent increase in new business start-up activity when voice and accountability is high (2.5). Similarly, our estimates suggest a ten percent increase in natural disaster activity is associated with a 0.5 percent decrease in new business start-up activity when political stability is low (-2.5) and a 0.36 percent increase in new business start-up activity when political stability is high (2.5).

---------------------------------
Insert Figure 2 About Here
---------------------------------

Figure 3 plots the moderating effects of the two governance indicators of government effectiveness and regulatory quality. The results indicate that natural disasters are negatively associated with new business start-up activity, but this effect becomes smaller or positive as the quality of government effectiveness and regulatory quality increase. Specifically, our estimates suggest a ten percent increase in natural disaster activity is associated with an one percent decrease in new business start-up activity when government effectiveness is low (-2.5) and a 0.82 percent increase in new business start-up activity when government effectiveness is high (2.5). Regarding the regulatory quality indicator, our estimates suggest that a ten percent increase in natural disaster activity is associated with a one percent decrease in new business start-up activity when regulatory quality is low (-2.5) and a 0.66 percent increase in new business start-up activity when regulatory quality is high (2.5).

---------------------------------
Insert Figure 3 About Here
---------------------------------

Figure 4 plots the moderating effects of the last two governance indicators, namely rule of law and control of corruption. The results of these indicators are very similar to the previous findings; that is, natural disasters are negatively associated with new business start-up activity, but this effect becomes smaller or positive as the rule of law score and control of corruption quality score increase. Our estimates suggest that a ten percent increase in natural disaster activity is associated with a 0.82 percent decrease in new business start-up activity when the rule of law is low (-2.5) and a 0.64 percent increase in new business start-up activity when the rule of law is high (2.5). Regarding the control of corruption indicator, our estimates suggest that a ten percent increase in natural disaster activity is associated with a one percent decrease in new business start-up activity when control of corruption is low (-2.5) and a 0.87 percent increase in new business start-up activity when control of corruption is high (2.5).

---------------------------------
Insert Figure 4 About Here
---------------------------------

### 4.4. Analytic Extensions
#### 4.4.1. Are these short-term or long-term effects?

While it is not the focus our study, a natural question that arises is whether the moderating role of a country's quality of governance persists for more than a year. We examine these short- and long-term effects by lagging our natural disaster variable by two, three, and four years. By including lags in the model, we are better able to examine the rate of start-up activity two, three, and four years after a natural disaster. For instance, a one-year lag suggests that natural disasters have an effect in the next year, not the same year. The same is true for two, three, or four years. Our results in Tables 6, 7, and 8 indicate that natural disasters' effects on start-up activity persist for the short term (1–3 years) but not the long term. We also found similar but weaker evidence regarding a four-year lag structure. Nevertheless, we found that country governance positively moderates the relationship between natural disasters and start-up activity for two years, three years, and possibly four years. These findings also indicate that the type of country governance matters: we found evidence for government effectiveness, regulatory quality, rule of law, and control of corruption. The evidence is weaker for voice and accountability and political stability.

---------------------------------------------
Insert Tables 6, 7, and 8 About Here
---------------------------------------------

#### 4.4.2. Do the findings differ by disaster type?

Another extension of our findings is to consider whether our findings differ by disaster type. That is, do all disasters have similar effects on start-up activity, or is it more heterogenous than that? To address this question, we separated disasters into climatic and geologic categories, following the literature (Boudreaux et al. 2019a; Crespo Cuaresma et al. 2008; Skidmore and Toya 2002). Climatic disasters include floods, cyclones, hurricanes, blizzards, typhoons, tornadoes, and storms. Geologic disasters include volcanic eruptions, natural explosions, landslides, avalanches, and earthquakes.

The results, which are available in a supplemental appendix (Tables A1 and A2), suggest that the moderating effect is statistically significant for geologic disasters but not climatic disasters. However, the results are likely more nuanced than that. For instance, a comparison of the moderating figures by disaster type suggests there is primarily a negative effect of climatic disasters on the rate of start-up activity (Figures A1-A3) yet primarily a positive effect of geologic disasters on the rate of start-up activity (Figures A4-A6). The marginal effect of climatic disasters does become positive, but it becomes positive for high-quality institutions only, typically at a score of 1.25 or greater. In contrast for geologic disasters, there is primarily a positive marginal effect of disasters on the rate of start-up activity, not negative. In almost an opposite fashion, the marginal effect is negative for low-quality institutions only. Thus, while we only observe a statistically significant moderating effect for geologic disasters (Table A2), we believe the findings are a bit more nuanced than that.

## 5. DISCUSSION AND CONCLUSIONS

The purpose of this paper is to investigate whether and how a country's quality of governance moderates the relationship between exogenous natural disasters and new business start-up activity. This study contributes to the literature by demonstrating that a country's quality of governance matters in the relationship between a natural disaster and the likelihood of a new business start-up activity. High-quality country-level governance makes a country more resilient to natural disasters. Our results suggest two possibilities: when a country's governance quality is high: 1) entrepreneurs expect the country to quickly recover from a natural disaster, and 2) entrepreneurs are in a better position to take advantage of new business opportunities in the aftermath of natural disasters.

This finding is an important contribution because although natural disasters are devastating and their frequency is increasing, we know little about what makes one country more resilient than others in dealing with these disasters. The following quotation illustrates this point:

> Why do some organizations and societies successfully adjust and even thrive amid adversity while others fail to do so? With this editorial, we would like to inspire management scholars to take up the "grand challenge" of studying the role and functioning of organizations during adverse natural or social events" (Van Der Vegt et al. 2015).

Recently, researchers have pointed to the lack of research on this topic (e.g., Williams and Shepherd 2016). By documenting that high-quality country-level governance is associated with more new business start-ups after a natural disaster, our research suggests that formal institutions such as government effectiveness, lower corruption, private sector-friendly regulations, the rule of law, and citizens' ability to voice their opinion freely not only help countries recover but also allow entrepreneurs to thrive in post-disaster periods. This is a fresh perspective. The few studies focusing on this line of research suggest informal institutions such as cultural traits like social capital—the norms and networks of a society that enable co-operation—play a major role in a country's recovery after a natural disaster (Aldrich 2012). However, the role of formal institutions has largely been ignored. Our study fills this gap by highlighting how formal institutions also play an important role in facilitating post-disaster recovery through start-up activity.

Birkmann et al. (2010) have theorized that extreme events such as disasters can create a window of opportunity for change; our study suggests that this may only be possible if the country's formal governance is strong. Otherwise, business activity might suffer. As we mentioned in the introduction, Oh and Oetzel (2011) have found that only one aspect of a formal institution—regulatory quality—positively moderates the effect of natural disasters on MNCs' subsidiary-level investment. In contrast, we found that *all* aspects of high-quality country-level governance positively moderate the effect of natural disasters on start-up activity. More specifically, based on Oh and Oetzel's study, government effectiveness, the rule of law, and the level of corruption do not affect MNCs' decisions to increase the number of foreign subsidiaries in the year after a disaster. However, as our research demonstrates, these aspects of country governance positively affect the number of new business start-ups in the year following a disaster.

Our results are consistent with the idea that while all businesses bear the cost of low-quality governance after a disaster, start-ups are disproportionately affected since they are often ill equipped to navigate poor governance when a natural disaster strikes. Oh and Oetzel (2011) have noted this possibility, "entrepreneurial firms are likely to have different decision-making rules and processes against disasters due to the differences in risk-taking behaviors and organizational responsiveness" (page 678). Our results confirm Oh and Oetzel's suspicion that the effect of institutions on moderating a natural disaster's effect on new business start-ups is even stronger.

Entrepreneurship researchers have found that institutional quality, especially the regulatory framework, positively affect the number of start-ups. The regulatory framework appears to affect replicative entrepreneurship more than high-impact entrepreneurship (Stenholm et al. 2013). Our study extends this stream of literature by demonstrating that regulatory quality has an impact on the number of start-ups and that all types of institutional quality affect entrepreneurship in the aftermath of a natural disaster. Therefore, our results suggest that the impact of institutions might be particularly valuable in the event of a negative shock to the economy.

Although we investigate the likelihood of starting a new business, the implication of our study is also to a related topic—the survival of small businesses. In this respect, we extend several recent pieces of research on entrepreneurship that examine the survival of small firms. Davlasheridze and Geylani (2017) find that small businesses are more likely to thrive if they have better loan access after being struck by natural disasters. Hadjielias et al. (2022) suggest that the psychological resilience of the small business owner increases the possibility of business survival during a disaster. Grözinger et al. (2022) echo a similar thought. They empirically document that organization psychological capital captured by solidarity, co-operation, and citizenship behavior of the employees is positively associated with innovation for survival during a disaster. We extend this study because our study suggests that higher-quality formal institutions might be another factor that can help the survival of small businesses thrive during a natural disaster.

Our study also has managerial implications. Churchill and Lewis (1983) have classified the five stages of new business start-up activity. They have theorized that in the first stage, entrepreneurs ask the following questions:

> "Can we get enough customers, deliver our products, and provide services well enough to become a viable business? Can we expand from that one key customer or pilot production process to a much broader sales base? Do we have enough money to cover the considerable cash demands of this start-up phase?" (page 3)

Our results indirectly suggest that in countries with high-quality governance, entrepreneurs should expect that the infrastructure will improve quickly; that they will be able to reach customers, provide services, and expand quickly; and that they can be confident about obtaining resources such as loans from banks to grow more easily.

From a policy perspective, our study highlights the importance of designing policies that strengthen a country's governance quality. Research demonstrates that high-quality institutions are positively associated with foreign direct investment (FDI) (Globerman and Shapiro 2003) and high-growth entrepreneurial activity (Bowen and De Clercq 2008). They can also better withstand the initial disaster shock on the GDP growth rate (Noy 2009; Raschky 2008). Our research shows high-quality country governance makes a country resilient to the adverse effect of a natural disaster on start-up activity.

### 5.1. Limitations and future research

One of our limitations is the unit of analysis. Our measures of new business start-up activity and country governance are all at the country–year level. Therefore, when we claim that supply chain networks are disrupted, we implicitly assume that disruption occurred in the entire country. However, this is not accurate, particularly if the country is large. For example, Hurricane Katrina caused severe damage in Louisiana, but the rest of the country was not affected. We believe that this is one of the limitations of our study, albeit one that is difficult to solve, because most variations in institutional quality are at the country level (i.e., country governance). In any case, an alternative way to address the question we investigated would be to use a single-country study with variation in institutional quality in different regions. This approach would be more precise in capturing the cost of natural disasters and the opportunities they create and could provide additional insight into our hypotheses. The downside of this strategy is that there is less heterogeneity in institutional quality *within* a country than there is *between* countries. Thus, there were certain benefits to broadening our perspective to conduct a multi-country study.

Due to data limitations, we were also unable to examine what types of new start-ups are likely to be moderated by the quality of governance. A dataset that provides information about the types of start-ups and the characteristics of the entrepreneurs could be used to better understand the moderating role of the quality of governance after a disaster.

Another limitation is that we have assumed that the quality of institutions is not affected by the natural disaster. While largely true in the short term, this assumption is inaccurate in the long term. When a natural disaster strikes, there is often an inflow of aid from the government, non-government organizations, and foreign countries, partly for altruistic reasons and partly for political reasons (Garrett and Sobel 2003; Sobel and Leeson 2006). This sudden inflow of capital from various sources can increase corruption. Leeson and Sobel (2008) posit that some parts of the US may be more corrupt because they are struck by natural disasters often. The idea is that disasters trigger federally funded relief, which increases the chance of theft, and makes the culture more corrupt over time. The inflow of cash from different sources and susceptibility to corruption may lead to more regulations in regions prone to natural disasters. However, the adverse impact of the disaster on the culture and regulations will be gradual. Nevertheless, examining how the level of corruption and regulations changes after a natural disaster and how these changes affect the subsequent likelihood of startups in the long term is a possible avenue for future research.

We also want to reiterate that our results indicate that quality institutions have a *short-term* increase in the number of new firms registered after a natural disaster. This limitation is a caveat. Bastiat (1964) famous broken window fallacy postulated that when a shopkeeper's son breaks a glass in his father's store, the society will be worse off, not better off, simply because the resource spent on replacing the glass could have been used elsewhere. The idea's extension to a natural disaster means that the country is worse off after a natural disaster. However, as Diaz and Larroulet (2021) point out, not all economists agree. For example, if a natural disaster leads to a significant innovation that considerably impacts the economy, one could argue that there may be long-term benefits. This harbinger of creative destruction predicts a higher growth rate and, thus, more new businesses getting established (Hallegatte and Dumas 2009). While the theory is plausible, the extent and the likelihood of long-term benefits due to a natural disaster are hard to assess and are not the focus of our study. Future research could examine these questions in greater depth. For example, such benefits may only incur in well-developed countries where access to credit and venture capital is more accessible.

By demonstrating that a country's quality of governance makes that country more resilient to natural disasters, our study raises new questions. For example, recent studies have shown that natural disaster risk can affect a firm's financing policies, such as the amount cash, the type of debt, and firm performance (Huang et al. 2018). Our study indirectly suggests that these effects of natural disasters may be moderated by the quality of country-level governance.

Another possible extension of our study would be to compare the moderating role of the quality of governance in the investment decisions of foreign and domestic firms after a natural disaster. The abilities of foreign and domestic firms to navigate the changing landscape after a disaster might differ, and good country governance might moderate such an effect.

Our research also raises similar questions in emerging strands of literature about natural disasters and economic growth. It is unclear exactly how natural disasters affect economic growth. Some economic models use Schumpeter's theory of creative destruction to argue that disasters can be a catalyst in upgrading destroyed capital stock and hence result in long-term economic growth (Crespo Cuaresma et al. 2008; Toya and Skidmore 2007). Toya and Skidmore (2007) have found that climatic disasters are positively correlated with economic growth. Others disagree, arguing that natural disaster can lead to permanent destruction of physical or human capital and negative deviation from a previous growth trajectory (Romer 1990). Felbermayr and Gröschl (2014) have used disaster data from geophysical and meteorological sources rather than insurance companies regarding more than 100 countries and spanning three decades. They found that the worst five percent of disaster years were characterized by growth damage of at least 0.46 percent. Bergholt and Lujala (2012) have also found that the negative effect of natural disasters on growth is considerable. Our study indirectly raises the possibility that natural disasters' impact on economic growth might be moderated by the quality of country governance factors such as the rule of law, corruption, voice and accountability, and regulatory quality.

# APPENDIX

**Table A1**. Robustness Check 1: Omitting GDP per capita does not affect results
Dependent variable: start-up rate

|  | Country Governance | | | | | |
|---|---|---|---|---|---|---|
|  | Voice and Accountability (1) | Political Stability (2) | Government Effectiveness (3) | Regulatory Quality (4) | Rule of Law (5) | Control of Corruption (6) |
| *Explanatory Variables* | | | | | | |
| Natural Disasters $_{t-1}$ | -0.030 | -0.020 | -0.023 | -0.025 | -0.020 | -0.018 |
|  | (0.001) | (0.024) | (0.009) | (0.006) | (0.023) | (0.029) |
| Country Governance $_{t-1}$ | 0.401 | 0.134 | 0.409 | 0.540 | 0.240 | 0.207 |
|  | (0.000) | (0.133) | (0.000) | (0.000) | (0.005) | (0.004) |
| *Interaction Terms* | | | | | | |
| Natural Disasters | 0.015 | 0.006 | 0.027 | 0.021 | 0.020 | 0.031 |
| X Country Governance $_{t-1}$ | (0.055) | (0.490) | (0.003) | (0.017) | (0.015) | (0.000) |
| *Control Variables* | | | | | | |
| Cost of start-up procedures | -0.004 | -0.004 | -0.004 | -0.003 | -0.004 | -0.004 |
|  | (0.013) | (0.012) | (0.019) | (0.023) | (0.015) | (0.016) |
| Time required to start a business | -0.006 | -0.006 | -0.003 | -0.004 | -0.005 | -0.005 |
|  | (0.001) | (0.001) | (0.105) | (0.025) | (0.010) | (0.007) |
| Number of start-up procedures | -0.064 | -0.071 | -0.066 | -0.056 | -0.067 | -0.061 |
|  | (0.000) | (0.000) | (0.000) | (0.000) | (0.000) | (0.000) |
| Largest population | 0.898 | 1.546 | 0.840 | 0.779 | 1.206 | 1.041 |
|  | (0.050) | (0.001) | (0.065) | (0.085) | (0.008) | (0.021) |
| GDP growth | 0.014 | 0.007 | 0.010 | 0.014 | 0.007 | 0.006 |
|  | (0.171) | (0.520) | (0.341) | (0.173) | (0.506) | (0.579) |
| Population density | -0.002 | -0.001 | -0.002 | -0.002 | -0.002 | -0.002 |
|  | (0.000) | (0.000) | (0.000) | (0.000) | (0.000) | (0.000) |
| Domestic credit | 0.004 | 0.008 | 0.003 | 0.003 | 0.005 | 0.005 |
|  | (0.000) | (0.000) | (0.003) | (0.005) | (0.000) | (0.000) |
| Land area | -0.010 | -0.022 | -0.070 | -0.029 | -0.051 | -0.052 |
|  | (0.804) | (0.563) | (0.083) | (0.455) | (0.202) | (0.193) |
| Trade | 0.002 | 0.001 | 0.001 | 0.001 | 0.001 | 0.002 |
|  | (0.038) | (0.302) | (0.624) | (0.467) | (0.203) | (0.033) |
| Observations | 703 | 703 | 703 | 703 | 703 | 703 |
| Adjusted $R^2$ | 0.544 | 0.503 | 0.543 | 0.556 | 0.519 | 0.528 |
| F-test | 61.556 | 47.415 | 57.437 | 65.702 | 54.117 | 59.155 |
| Log-likelihood | -940.173 | -970.752 | -941.533 | -930.931 | -959.315 | -952.573 |

Note – P-value in parentheses (two-tailed test). Standard errors are robust to heteroscedasticity. Year fixed effects and constant included in all models but omitted from table.

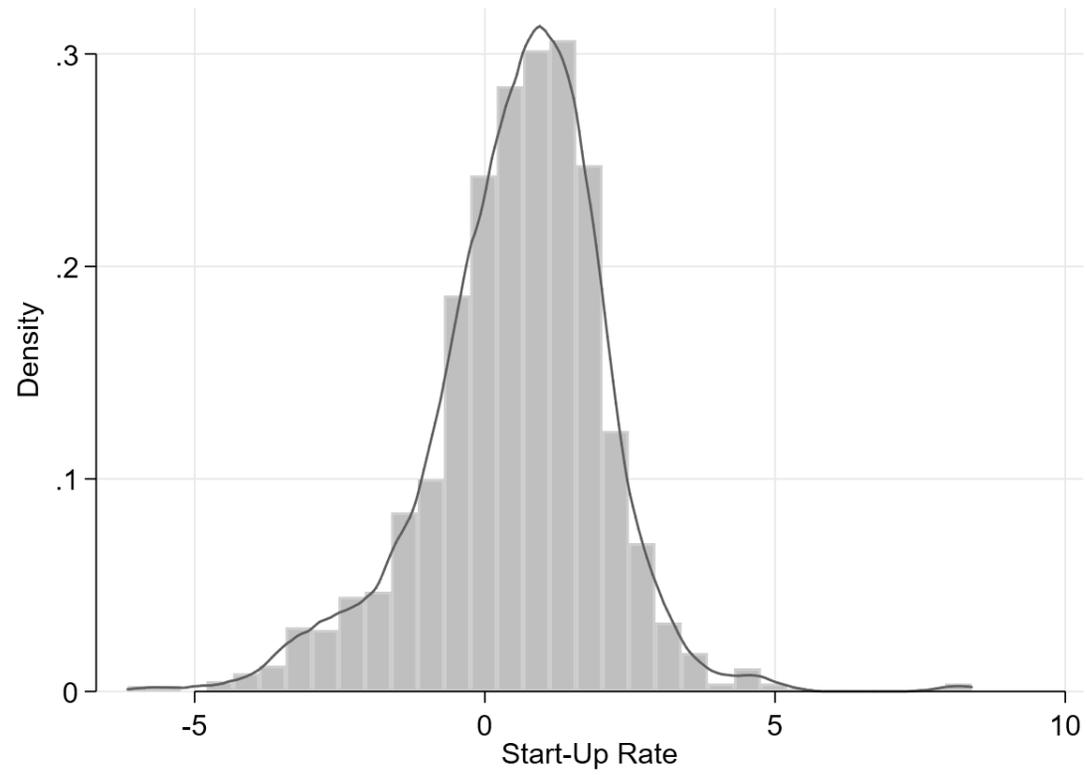

**Figure 1**.
The Dependent Variable, Start-up Rate, is normally distributed after a log-transformation

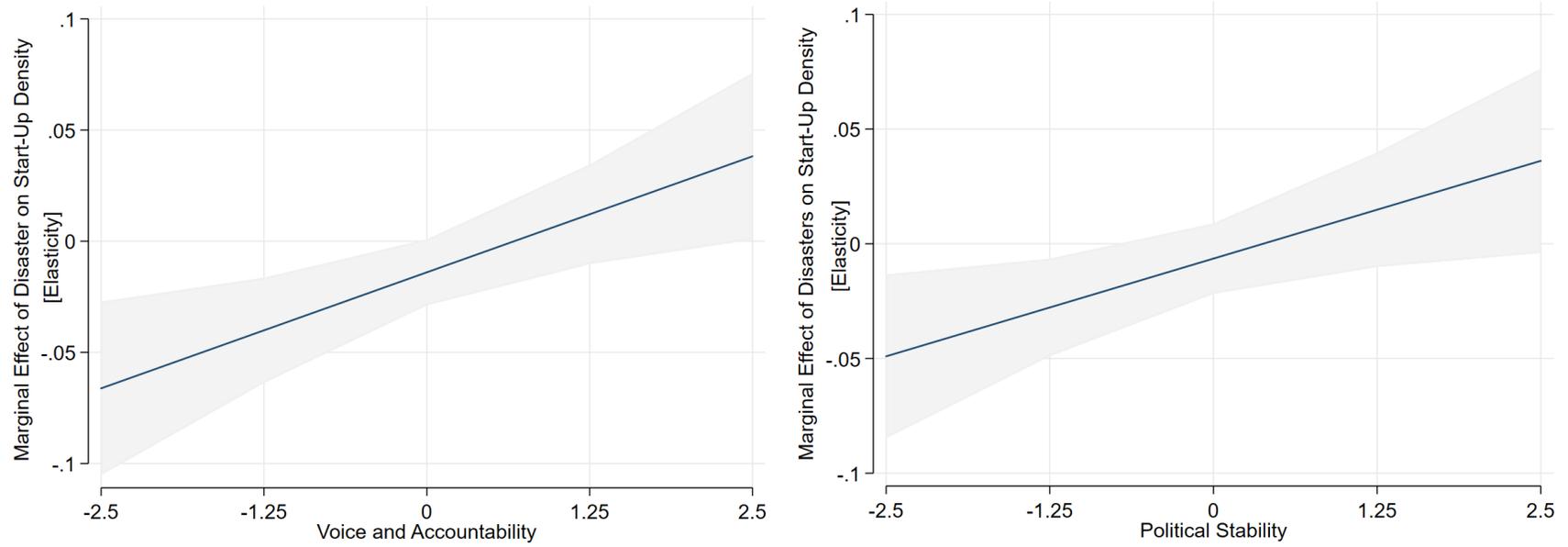

**Figure 2.**
Effect of Disasters on Start-Up Rate is Moderated by Voice and Accountability and Political Stability [95% confidence intervals]

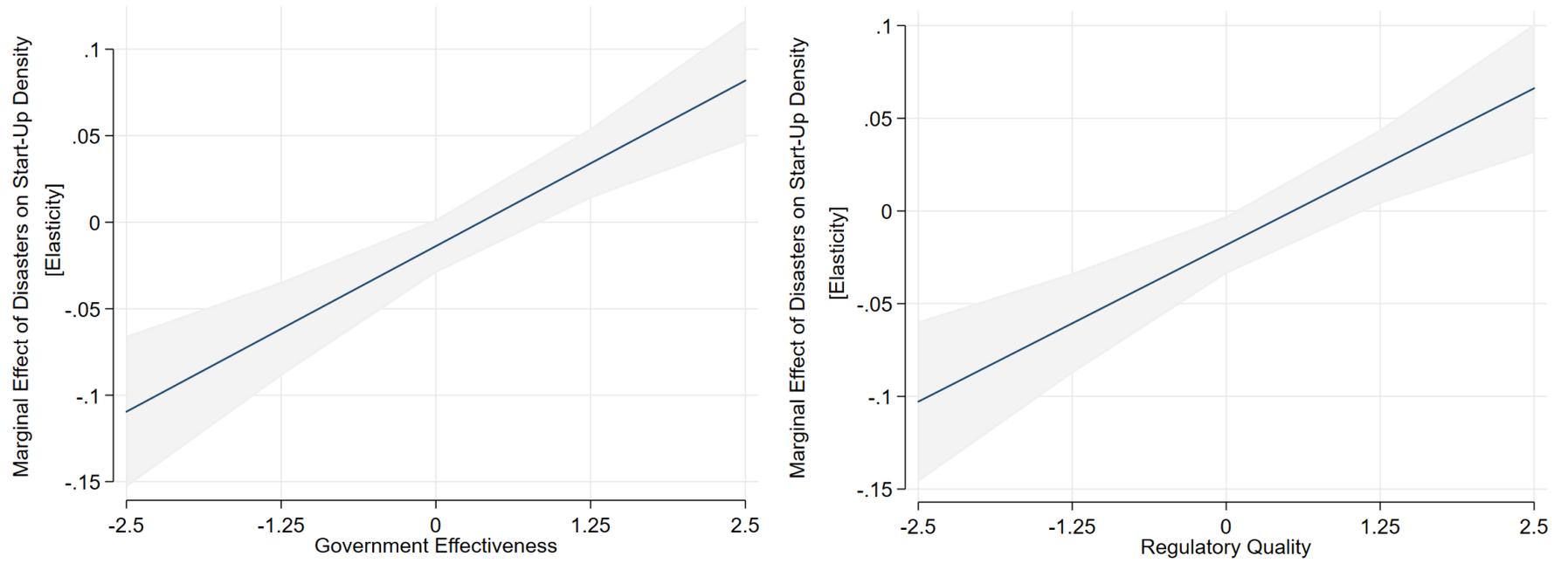

**Figure 3.**
Effect of Disasters on Start-Up Rate is Moderated by Government Effectiveness and Regulatory Quality [95% confidence intervals]

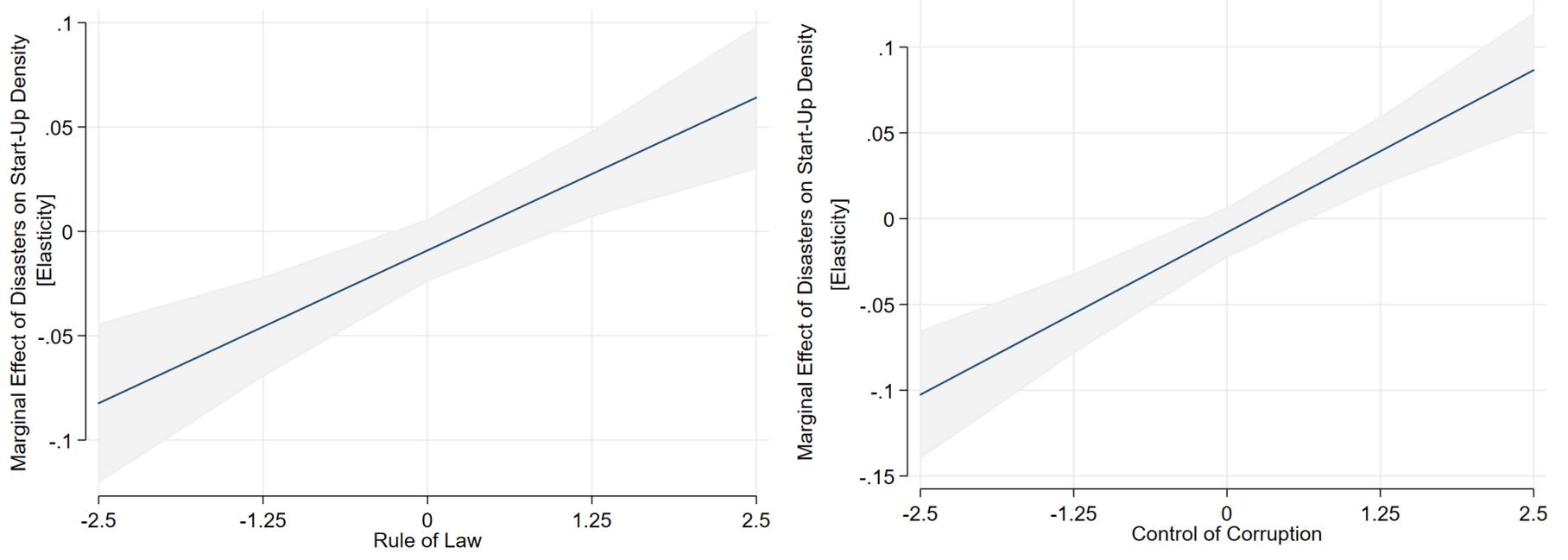

**Figure 4.**
Effect of Disasters on Start-Up Rate is Moderated by Rule of Law and Control of Corruption [95% confidence intervals]

**Table 1.** Definitions and data sources of variables.

| Variable | Definition | Data source |
|---|---|---|
| *Dependent variable* | | |
| Start-up rate | Log of new business registrations per 1000 people ages 15-64. | World Bank's Entrepreneurship Survey and Database |
| *Explanatory variables* | | |
| Natural Disasters | Log of (1 + total number of people affected, injured, and made homeless by natural disasters). | EM-DAT |
| Voice and accountability | Score measure of the ability of citizens' participation in the government. | WGI |
| Political stability | Score measure of the ability of the government to destabilize violence and terrorism. | WGI |
| Government effectiveness | Score measure of the quality of public service and its independence, policy formulation and implementation, and credibility of government policies. | WGI |
| Regulatory quality | Score measure of the ability of the government to formulate and implement sound policies and regulations. | WGI |
| Rule of law | Score measure of the quality of contract enforcement, property rights, the police, and the courts. | WGI |
| Control of corruption | Score measure of the extent to which public power is exercised for private gain. | WGI |
| *Control variables* | | |
| GDP per capita | Log of Real GDP per capita using purchasing power parity for international comparisons. | |
| Cost of business start-up procedures | Cost of business start-up procedures as a percent of GNI per capita. | World Bank, Doing Business |
| Time required to start a business | Number of days required to start a business. | World Bank, Doing Business |
| Number of start-up procedures | Number of start-up procedures to register a business. | World Bank, Doing Business |
| Largest population | Largest city population as a percent of total population. | World Bank Open Data |
| GDP growth | Annual growth rate of GDP. | World Bank Open Data |
| Population density | People per square kilometer measured as population divided by land area. | World Bank Open Data |
| Domestic credit | Domestic credit provided by financial sector as a percent of GDP. | World Bank Open Data |
| Land area | Log of land area in square kilometers. | World Bank Open Data |
| Trade | Sum of exports and imports of goods and services measured as a percent of GDP. | World Bank Open Data |

**Table 2.** Summary Statistics by Country (2006-2016)

| Country | Start-up Rate | Disaster Intensity | Voice and Accountability | Political Stability | Government Effectiveness | Regulatory Quality | Rule of Law | Control of Corruption |
|---|---|---|---|---|---|---|---|---|
| Afghanistan | 0.47 | 109915 | -1.22 | -2.50 | -1.41 | -1.45 | -1.75 | -1.50 |
| Albania | 8.65 | 7494 | 0.11 | -0.06 | -0.26 | 0.16 | -0.50 | -0.62 |
| Algeria | 0.45 | 15572 | -0.96 | -1.21 | -0.54 | -1.00 | -0.77 | -0.55 |
| Argentina | 0.53 | 37140 | 0.35 | -0.01 | -0.15 | -0.80 | -0.65 | -0.41 |
| Armenia | 1.46 | 0 | -0.72 | -0.03 | -0.17 | 0.29 | -0.44 | -0.64 |
| Australia | 11.54 | 36760 | 1.40 | 0.94 | 1.70 | 1.77 | 1.79 | 1.98 |
| Austria | 0.63 | 57.33 | 1.39 | 1.22 | 1.69 | 1.52 | 1.87 | 1.65 |
| Azerbaijan | 0.76 | 13928 | -1.36 | -0.48 | -0.63 | -0.34 | -0.81 | -1.12 |
| Belarus | 0.82 | 0 | -1.59 | 0.19 | -0.95 | -1.20 | -1.04 | -0.56 |
| Belgium | 3.04 | 84.78 | 1.35 | 0.78 | 1.55 | 1.30 | 1.39 | 1.48 |
| Bolivia | 0.46 | 189326 | -0.01 | -0.53 | -0.57 | -0.86 | -1.00 | -0.56 |
| Brazil | 0.50 | 463850 | 0.47 | -0.20 | -0.16 | -0.14 | -0.10 | -0.37 |
| Bulgaria | 8.14 | 6685 | 0.47 | 0.27 | 0.11 | 0.61 | -0.08 | -0.23 |
| Burkina Faso | 0.11 | 63698 | -0.29 | -0.09 | -0.62 | -0.20 | -0.39 | -0.35 |
| Chile | 4.81 | 488847 | 1.06 | 0.49 | 1.20 | 1.47 | 1.34 | 1.44 |
| Colombia | 1.59 | 871941 | -0.11 | -1.53 | -0.07 | 0.30 | -0.36 | -0.29 |
| Costa Rica | 1.71 | 52141 | 1.03 | 0.60 | 0.31 | 0.47 | 0.49 | 0.64 |
| Croatia | 3.31 | 1346 | 0.51 | 0.60 | 0.59 | 0.48 | 0.17 | 0.08 |
| Czech Republic | 3.01 | 2262 | 1.01 | 1.02 | 0.97 | 1.14 | 1.00 | 0.35 |
| Denmark | 7.48 | 0 | 1.55 | 1.02 | 2.10 | 1.83 | 1.97 | 2.37 |
| Dom. Republic | 0.93 | 41598 | 0.14 | 0.04 | -0.56 | -0.15 | -0.65 | -0.79 |
| El Salvador | 0.50 | 49088 | 0.08 | 0.02 | -0.12 | 0.27 | -0.66 | -0.34 |
| Estonia | 14.62 | 0 | 1.11 | 0.65 | 1.06 | .145 | 1.21 | 1.08 |
| Finland | 3.55 | 0 | 1.51 | 1.38 | 2.10 | 1.76 | 1.98 | 2.28 |
| France | 2.96 | 56815 | 1.24 | 0.50 | 1.46 | 1.20 | 1.47 | 1.43 |
| Gabon | 4.04 | 200 | -0.91 | 0.23 | -0.78 | -0.54 | -0.65 | -1.03 |
| Georgia | 4.06 | 15335 | -0.06 | -0.70 | 0.31 | 0.55 | -0.13 | 0.19 |
| Germany | 1.28 | 48577 | 1.36 | 0.87 | 1.62 | 1.59 | 1.72 | 1.79 |
| Ghana | 0.94 | 94480 | 0.47 | 0.04 | 0.02 | 0.04 | -0.01 | -0.03 |
| Greece | 0.92 | 1612 | 0.93 | 0.22 | 0.59 | 0.82 | 0.78 | 0.15 |
| Guatemala | 0.68 | 201138 | -0.23 | -0.79 | -0.64 | -0.18 | -1.06 | -0.63 |
| Guinea | 0.13 | 0 | -0.86 | -0.96 | -1.23 | -1.06 | -1.36 | -1.06 |
| Haiti | 0.05 | 862853 | -0.66 | -1.20 | -1.40 | -0.88 | -1.34 | -1.21 |
| Hungary | 5.02 | 4533 | 0.84 | 0.73 | 0.66 | 1.03 | 0.75 | 0.40 |
| India | 0.09 | 11200000 | 0.42 | -0.97 | -0.06 | -0.42 | -0.06 | -0.39 |
| Indonesia | 0.34 | 388960 | 0.17 | -0.52 | -0.14 | -0.16 | -0.38 | -0.51 |
| Iraq | 2.66 | 14501 | -1.11 | -2.33 | -1.28 | -1.19 | -1.58 | -1.34 |
| Ireland | 5.31 | 67 | 1.35 | 1.06 | 1.49 | 1.74 | 1.75 | 1.66 |
| Israel | 3.29 | 22 | 0.68 | -1.25 | 1.30 | 1.17 | 0.95 | 0.87 |
| Italy | 2.21 | 10963 | 1.01 | 0.47 | 0.37 | 0.84 | 0.41 | 0.18 |
| Jamaica | 1.16 | 28949 | 0.56 | -0.17 | 0.20 | 0.24 | -0.36 | -0.26 |
| Japan | 0.15 | 49856 | 1.04 | 0.97 | 1.81 | 1.14 | 1.60 | 1.70 |
| Jordan | 0.77 | 0 | -0.75 | -0.48 | 0.15 | 0.23 | 0.34 | 0.18 |
| Kazakhstan | 1.85 | 7775 | -1.14 | 0.23 | -0.36 | -0.29 | -0.70 | -0.94 |
| Kenya | 0.68 | 296138 | -0.22 | -1.27 | -0.57 | -0.22 | -0.95 | -0.99 |
| Korea, Rep. | 1.81 | 9107 | 0.70 | 0.34 | 1.14 | 0.92 | 0.97 | 0.49 |
| Kyrgyz Republic | 0.97 | 6922 | -0.77 | -0.92 | -0.78 | -0.39 | -1.20 | -1.22 |
| Lao PDR | 0.18 | 111026 | -1.70 | -0.14 | -0.89 | -1.08 | -0.97 | -1.21 |
| Latvia | 10.38 | 0 | 0.80 | 0.45 | 0.86 | 1.04 | 0.80 | 0.33 |

| Country | Start-up rate | Disaster Intensity | Voice and Accountability | Political Stability | Government Effectiveness | Regulatory Quality | Rule of Law | Control of Corruption |
|---|---|---|---|---|---|---|---|---|
| Lithuania | 3.85 | 0 | 0.93 | 0.73 | 0.89 | 1.10 | 0.87 | 0.45 |
| Madagascar | 0.10 | 104087 | -0.44 | -0.50 | -1.30 | -0.73 | -0.71 | -0.84 |
| Malawi | 0.09 | 17709 | -0.23 | 0.05 | -0.60 | -0.47 | -0.15 | -0.48 |
| Malaysia | 2.31 | 59683 | -0.47 | 0.14 | 1.08 | 0.56 | 0.47 | 0.15 |
| Mexico | 0.55 | 96654 | -0.04 | -0.83 | 0.20 | 0.39 | -0.43 | -0.76 |
| Moldova | 1.90 | 1778 | -0.13 | -0.23 | -0.65 | -0.14 | -0.39 | -0.70 |
| Mongolia | 5.25 | 1250 | 0.18 | 0.62 | -0.56 | -0.27 | -0.33 | -0.63 |
| Morocco | 1.34 | 25070 | -0.69 | -0.44 | -0.12 | -0.17 | -0.21 | -0.33 |
| Myanmar | 0.15 | 0.30 | -1.39 | -1.07 | -1.33 | -1.48 | -1.26 | -0.93 |
| Namibia | 0.74 | 115883 | 0.44 | 0.89 | 0.12 | 0.05 | 0.18 | 0.34 |
| Nepal | 0.62 | 792428 | -0.53 | -1.48 | -0.89 | -0.70 | -0.74 | -0.72 |
| Netherlands | 3.92 | 0 | 1.52 | 0.97 | 1.77 | 1.76 | 1.84 | 2.08 |
| New Zealand | 16.94 | 60300 | 1.49 | 1.19 | 1.72 | 1.77 | 1.86 | 2.34 |
| Niger | 0.004 | 53719 | -0.47 | -0.63 | -0.73 | -0.48 | -0.58 | -0.73 |
| Nigeria | 0.75 | 55210 | -0.48 | -2.03 | -1.07 | -0.83 | -1.01 | -1.18 |
| Norway | 6.71 | 66.67 | 1.61 | 1.25 | 1.88 | 1.48 | 1.95 | 2.11 |
| Oman | 1.08 | 2911 | -1.07 | 0.73 | 0.30 | 0.55 | 0.47 | 0.31 |
| Pakistan | 0.05 | 2301965 | -0.72 | -2.44 | -0.71 | -0.65 | -0.76 | -0.82 |
| Panama | 0.56 | 9183 | 0.56 | 0.01 | 0.17 | 0.40 | -0.11 | -0.30 |
| Paraguay | 0.19 | 61016 | -0.12 | -0.62 | -0.91 | -0.43 | -0.85 | -0.92 |
| Peru | 2.97 | 347063 | 0.11 | -0.91 | -0.39 | 0.34 | -0.65 | -0.23 |
| Philippines | 0.27 | 8777178 | -0.02 | -1.39 | 0.05 | -0.12 | -0.47 | -0.66 |
| Poland | 0.77 | 11558 | 0.99 | 0.86 | 0.59 | 0.91 | 0.65 | 0.50 |
| Portugal | 4.44 | 114 | 1.14 | 0.83 | 1.02 | 0.90 | 1.06 | 1.03 |
| Russia | 4.36 | 13023 | -1.07 | -0.99 | -0.16 | -0.45 | -0.75 | -0.93 |
| Rwanda | 0.77 | 4736 | -1.25 | -0.32 | -0.09 | -0.23 | -0.34 | 0.29 |
| Saudi Arabia | 0.34 | 1121 | -1.81 | -0.43 | -0.04 | 0.06 | 0.09 | -0.06 |
| Senegal | 0.31 | 41916 | -0.06 | -0.22 | -0.43 | -0.25 | -0.28 | -0.35 |
| Serbia | 1.84 | 17535 | 0.27 | -0.26 | -0.06 | -0.07 | -0.33 | -0.30 |
| Sierra Leone | 0.22 | 4050 | -0.24 | -0.20 | -1.20 | -0.87 | -0.91 | -0.90 |
| Slovak Republic | 4.04 | 378 | 0.94 | 0.98 | 0.84 | 1.01 | 0.54 | 0.27 |
| South Africa | 8.00 | 46766 | 0.61 | -0.02 | 0.41 | 0.42 | 0.14 | 0.13 |
| Spain | 3.19 | 2549 | 1.10 | -0.12 | 1.02 | 1.06 | 1.11 | 1.01 |
| Sri Lanka | 0.39 | 602992 | -0.50 | -1.21 | -0.14 | -0.23 | -0.06 | -0.30 |
| Sweden | 5.64 | 0 | 1.56 | 1.14 | 1.93 | 1.71 | 1.95 | 2.23 |
| Switzerland | 4.22 | 345 | 1.56 | 1.31 | 1.98 | 1.63 | 1.83 | 2.11 |
| Tajikistan | 0.29 | 14112 | -1.40 | -0.99 | -0.98 | -1.05 | -1.19 | -1.17 |
| Thailand | 0.71 | 0.28 | -0.59 | -1.18 | 0.29 | 0.22 | -0.16 | -0.38 |
| Togo | 0.12 | 37462 | -0.98 | -0.26 | -1.41 | -0.90 | -0.89 | -0.99 |
| Tunisia | 1.34 | 744 | -1.05 | -0.09 | 0.29 | -0.03 | 0.03 | -0.19 |
| Turkey | 1.06 | 17237 | -0.15 | -0.99 | 0.29 | 0.32 | 0.04 | 0.03 |
| Uganda | 0.61 | 188191 | -0.49 | -0.98 | -0.55 | -0.22 | -0.36 | -0.84 |
| Ukraine | 1.00 | 30815 | -0.05 | -0.47 | -0.65 | -0.56 | -0.78 | -0.95 |
| United Kingdom | 11.16 | 52001 | 1.32 | 0.44 | 1.61 | 1.75 | 1.76 | 1.71 |
| Uruguay | 2.97 | 18925 | 1.09 | 0.87 | 0.52 | 0.39 | 0.66 | 1.28 |
| Zambia | 1.05 | 250412 | -0.18 | 0.42 | -0.68 | -0.49 | -0.41 | -0.41 |

**Table 3.** Descriptive statistics and correlation matrix

|   | Variables | Mean | S.D. | 1 | 2 | 3 | 4 | 5 | 6 | 7 | 8 | 9 | 10 | 11 | 12 | 13 | 14 | 15 | 16 | 17 | 18 |
|---|---|---|---|---|---|---|---|---|---|---|---|---|---|---|---|---|---|---|---|---|---|
| 1 | New business start-up rate | 0.28 | 1.37 | 1 | | | | | | | | | | | | | | | | | |
| 2 | Natural disasters | 5.56 | 5.31 | -0.38 | 1 | | | | | | | | | | | | | | | | |
| 3 | Voice and accountability | 0.19 | 0.93 | 0.50 | -0.20 | 1 | | | | | | | | | | | | | | | |
| 4 | Political violence | -0.02 | 0.89 | 0.51 | -0.32 | 0.67 | 1 | | | | | | | | | | | | | | |
| 5 | Government effectiveness | 0.25 | 0.96 | 0.58 | -0.19 | 0.75 | 0.69 | 1 | | | | | | | | | | | | | |
| 6 | Regulatory quality | 0.33 | 0.89 | 0.62 | -0.18 | 0.77 | 0.65 | 0.94 | 1 | | | | | | | | | | | | |
| 7 | Rule of law | 0.17 | 1.00 | 0.56 | -0.25 | 0.82 | 0.78 | 0.93 | 0.90 | 1 | | | | | | | | | | | |
| 8 | Control of corruption | 0.15 | 1.03 | 0.54 | -0.24 | 0.77 | 0.74 | 0.93 | 0.87 | 0.94 | 1 | | | | | | | | | | |
| 9 | GDP per capita | 9.40 | 1.10 | 0.61 | -0.20 | 0.48 | 0.55 | 0.77 | 0.74 | 0.70 | 0.69 | 1 | | | | | | | | | |
| 10 | Cost of business start-up procedures | 28.33 | 78.13 | -0.41 | 0.07 | -0.29 | -0.28 | -0.40 | -0.38 | -0.38 | -0.34 | -0.49 | 1 | | | | | | | | |
| 11 | Time required to start a business | 23.21 | 23.63 | -0.17 | 0.04 | -0.17 | -0.11 | -0.24 | -0.27 | -0.25 | -0.22 | -0.16 | 0.29 | 1 | | | | | | | |
| 12 | Number of start-up procedures | 7.75 | 3.29 | -0.40 | 0.23 | -0.42 | -0.34 | -0.44 | -0.45 | -0.47 | -0.44 | -0.27 | 0.32 | 0.50 | 1 | | | | | | |
| 13 | Largest population | 0.18 | 0.10 | 0.35 | -0.29 | 0.15 | 0.31 | 0.29 | 0.32 | 0.28 | 0.30 | 0.41 | -0.10 | -0.10 | -0.19 | 1 | | | | | |
| 14 | GDP growth | 3.62 | 4.10 | -0.17 | 0.10 | -0.21 | -0.12 | -0.15 | -0.16 | -0.18 | -0.17 | -0.13 | 0.09 | 0.10 | 0.16 | 0.01 | 1 | | | | |
| 15 | Population density | 114.4 | 115.4 | 0.13 | -0.01 | 0.08 | 0.15 | 0.20 | 0.21 | 0.16 | 0.15 | 0.19 | -0.07 | -0.09 | -0.16 | 0.62 | 0.05 | 1 | | | |
| 16 | Domestic credit | 77.16 | 60.52 | 0.44 | -0.08 | 0.55 | 0.38 | 0.69 | 0.64 | 0.67 | 0.65 | 0.54 | -0.29 | -0.24 | -0.42 | 0.07 | -0.21 | -0.01 | 1 | | |
| 17 | Land area | 12.22 | 1.41 | -0.13 | 0.42 | -0.35 | -0.48 | -0.22 | -0.22 | -0.35 | -0.29 | -0.12 | 0.10 | 0.06 | 0.15 | -0.59 | 0.12 | -0.34 | 0.08 | 1 | |
| 18 | Trade | 86.67 | 35.72 | 0.39 | -0.29 | 0.16 | 0.33 | 0.27 | 0.27 | 0.27 | 0.26 | 0.28 | -0.12 | -0.07 | -0.20 | 0.55 | -0.02 | 0.29 | 0.12 | -0.47 | 1 |

**Table 4.** Disasters, country governance, and start-up rate
Dependent variable: new business start-up rate

|  | Base model | Disasters | Country Governance | | | | | |
|---|---|---|---|---|---|---|---|---|
|  |  |  | Voice and Accountability | Political Stability | Government Effectiveness | Regulatory Quality | Rule of Law | Control of Corruption |
|  | (a) | (b) | (1) | (2) | (3) | (4) | (5) | (6) |
| *Explanatory Variables* | | | | | | | | |
| Natural Disasters $_{t-1}$ |  | -0.008 | -0.007 | -0.011 | -0.008 | -0.010 | -0.008 | -0.008 |
|  |  | (0.361) | (0.356) | (0.211) | (0.351) | (0.232) | (0.358) | (0.367) |
| Country Governance $_{t-1}$ |  |  | 0.351 | -0.152 | 0.048 | 0.269 | -0.077 | 0.021 |
|  |  |  | (0.000) | (0.027) | (0.628) | (0.008) | (0.414) | (0.780) |
| *Control Variables* | | | | | | | | |
| GDP per capita | 0.599 | 0.591 | 0.435 | 0.682 | 0.563 | 0.455 | 0.639 | 0.579 |
|  | (0.000) | (0.000) | (0.000) | (0.000) | (0.000) | (0.000) | (0.000) | (0.000) |
| Cost of business start-up procedures | -0.003 | -0.003 | -0.003 | -0.003 | -0.003 | -0.003 | -0.003 | -0.003 |
|  | (0.015) | (0.015) | (0.013) | (0.014) | (0.015) | (0.021) | (0.014) | (0.015) |
| Time required to start a business | -0.003 | -0.003 | -0.003 | -0.003 | -0.003 | -0.003 | -0.003 | -0.003 |
|  | (0.059) | (0.060) | (0.094) | (0.065) | (0.062) | (0.092) | (0.053) | (0.061) |
| Number of start-up procedures | -0.088 | -0.087 | -0.079 | -0.091 | -0.086 | -0.075 | -0.090 | -0.086 |
|  | (0.000) | (0.000) | (0.000) | (0.000) | (0.000) | (0.000) | (0.000) | (0.000) |
| Largest population | 0.389 | 0.313 | 0.278 | 0.164 | 0.326 | 0.323 | 0.333 | 0.313 |
|  | (0.446) | (0.545) | (0.573) | (0.761) | (0.526) | (0.524) | (0.524) | (0.545) |
| GDP growth | 0.022 | 0.021 | 0.024 | 0.022 | 0.021 | 0.020 | 0.022 | 0.021 |
|  | (0.024) | (0.025) | (0.011) | (0.019) | (0.028) | (0.041) | (0.023) | (0.029) |
| Population density | -0.002 | -0.002 | -0.002 | -0.002 | -0.002 | -0.002 | -0.002 | -0.002 |
|  | (0.000) | (0.000) | (0.000) | (0.000) | (0.000) | (0.000) | (0.000) | (0.000) |
| Domestic credit | 0.003 | 0.003 | 0.001 | 0.003 | 0.002 | 0.001 | 0.003 | 0.003 |
|  | (0.007) | (0.007) | (0.526) | (0.003) | (0.030) | (0.202) | (0.008) | (0.028) |
| Land area | -0.172 | -0.166 | -0.136 | -0.192 | -0.161 | -0.134 | -0.173 | -0.164 |
|  | (0.000) | (0.000) | (0.002) | (0.000) | (0.000) | (0.005) | (0.000) | (0.000) |
| Trade | -0.003 | -0.003 | -0.002 | -0.003 | -0.003 | -0.003 | -0.003 | -0.003 |
|  | (0.018) | (0.014) | (0.027) | (0.024) | (0.016) | (0.019) | (0.013) | (0.019) |
| Observations | 701 | 701 | 701 | 701 | 701 | 701 | 701 | 701 |
| Adjusted $R^2$ | 0.582 | 0.583 | 0.599 | 0.582 | 0.583 | 0.591 | 0.582 | 0.582 |
| F-test | 59.093 | 57.256 | 63.323 | 54.245 | 56.129 | 63.521 | 55.687 | 56.872 |
| Log-likelihood | -899.588 | -898.962 | -884.426 | -898.917 | -898.035 | -891.271 | -898.912 | -898.814 |

Note – P-values in parentheses (two-tailed test). Standard errors are robust to heteroscedasticity. Year fixed effects, regional fixed effects, and constant included in all models but omitted from table.

**Table 5**. Effects of interaction between disasters and country governance
Dependent variable: New business start-up rate

| | Country Governance | | | | | |
|---|---|---|---|---|---|---|
| | Voice and Accountability (1) | Political Stability (2) | Government Effectiveness (3) | Regulatory Quality (4) | Rule of Law (5) | Control of Corruption (6) |
| *Explanatory Variables* | | | | | | |
| Natural Disasters $_{t-1}$ | -0.008 | -0.009 | -0.009 | -0.013 | -0.007 | -0.006 |
| | (0.339) | (0.302) | (0.319) | (0.175) | (0.445) | (0.460) |
| Country Governance $_{t-1}$ | 0.284 | -0.260 | -0.011 | 0.169 | -0.133 | -0.060 |
| | (0.000) | (0.005) | (0.915) | (0.133) | (0.194) | (0.459) |
| *Interaction Terms* | | | | | | |
| Natural Disasters | 0.014 | 0.014 | 0.020 | 0.021 | 0.014 | 0.025 |
| X Country Governance $_{t-1}$ | (0.064) | (0.063) | (0.017) | (0.012) | (0.058) | (0.001) |
| *Control Variables* | | | | | | |
| GDP per capita | 0.429 | 0.705 | 0.539 | 0.448 | 0.628 | 0.553 |
| | (0.000) | (0.000) | (0.000) | (0.000) | (0.000) | (0.000) |
| Cost of start-up procedures | -0.003 | -0.003 | -0.003 | -0.003 | -0.003 | -0.003 |
| | (0.016) | (0.016) | (0.019) | (0.024) | (0.016) | (0.018) |
| Time required to start a business | -0.002 | -0.003 | -0.002 | -0.002 | -0.003 | -0.003 |
| | (0.167) | (0.072) | (0.177) | (0.177) | (0.086) | (0.124) |
| Number of start-up procedures | -0.081 | -0.090 | -0.086 | -0.075 | -0.090 | -0.084 |
| | (0.000) | (0.000) | (0.000) | (0.000) | (0.000) | (0.000) |
| Largest population | 0.193 | 0.087 | 0.280 | 0.300 | 0.293 | 0.238 |
| | (0.697) | (0.872) | (0.582) | (0.551) | (0.573) | (0.641) |
| GDP growth | 0.024 | 0.022 | 0.022 | 0.022 | 0.023 | 0.022 |
| | (0.010) | (0.020) | (0.022) | (0.021) | (0.020) | (0.024) |
| Population density | -0.002 | -0.002 | -0.002 | -0.002 | -0.002 | -0.002 |
| | (0.000) | (0.000) | (0.000) | (0.000) | (0.000) | (0.000) |
| Domestic credit | 0.001 | 0.003 | 0.002 | 0.002 | 0.003 | 0.003 |
| | (0.445) | (0.002) | (0.033) | (0.149) | (0.007) | (0.027) |
| Land area | -0.145 | -0.197 | -0.166 | -0.143 | -0.176 | -0.167 |
| | (0.002) | (0.000) | (0.000) | (0.003) | (0.000) | (0.000) |
| Trade | -0.002 | -0.002 | -0.003 | -0.002 | -0.003 | -0.002 |
| | (0.032) | (0.049) | (0.019) | (0.026) | (0.021) | (0.042) |
| Observations | 701 | 701 | 701 | 701 | 701 | 701 |
| Adjusted $R^2$ | 0.603 | 0.585 | 0.599 | 0.602 | 0.592 | 0.600 |
| F-test | 63.130 | 53.083 | 57.030 | 63.944 | 55.266 | 58.835 |
| Log-likelihood | -880.642 | -896.122 | -884.229 | -881.582 | -890.105 | -883.155 |

Note – P-value in parentheses (two-tailed test). Standard errors robust to heteroscedasticity. Year fixed effects, regional fixed effects, and constant included in all models but omitted from table.

**Table 6.** Effects of interaction between disasters and country governance. **2-year lag**
Dependent variable: New business start-up rate

|  | Country Governance | | | | | |
|---|---|---|---|---|---|---|
|  | Voice and Accountability (1) | Political Stability (2) | Government Effectiveness (3) | Regulatory Quality (4) | Rule of Law (5) | Control of Corruption (6) |
| *Explanatory Variables* | | | | | | |
| Natural Disasters $_{t-2}$ | -0.005 | 0.001 | -0.003 | -0.006 | -0.000 | 0.001 |
|  | (0.549) | (0.934) | (0.723) | (0.534) | (0.966) | (0.934) |
| Country Governance $_{t-2}$ | 0.229 | -0.103 | 0.008 | 0.170 | -0.098 | -0.068 |
|  | (0.001) | (0.180) | (0.934) | (0.113) | (0.291) | (0.383) |
| *Interaction Terms* | | | | | | |
| Natural Disasters | 0.008 | 0.011 | 0.022 | 0.019 | 0.017 | 0.025 |
| X Country Governance $_{t-2}$ | (0.292) | (0.137) | (0.005) | (0.017) | (0.023) | (0.001) |
| *Control Variables* | | | | | | |
| GDP per capita | 0.507 | 0.606 | 0.543 | 0.483 | 0.599 | 0.571 |
|  | (0.000) | (0.000) | (0.000) | (0.000) | (0.000) | (0.000) |
| Cost of start-up procedures | -0.003 | -0.003 | -0.003 | -0.003 | -0.003 | -0.003 |
|  | (0.016) | (0.018) | (0.017) | (0.020) | (0.017) | (0.017) |
| Time required to start a business | -0.004 | -0.003 | -0.002 | -0.003 | -0.003 | -0.003 |
|  | (0.034) | (0.055) | (0.159) | (0.100) | (0.068) | (0.085) |
| Number of start-up procedures | -0.085 | -0.096 | -0.094 | -0.085 | -0.096 | -0.093 |
|  | (0.000) | (0.000) | (0.000) | (0.000) | (0.000) | (0.000) |
| Largest population | -0.099 | 0.009 | -0.060 | -0.024 | 0.010 | -0.061 |
|  | (0.830) | (0.985) | (0.900) | (0.958) | (0.983) | (0.898) |
| GDP growth | 0.023 | 0.020 | 0.022 | 0.023 | 0.021 | 0.021 |
|  | (0.015) | (0.035) | (0.026) | (0.017) | (0.029) | (0.031) |
| Population density | -0.002 | -0.002 | -0.002 | -0.002 | -0.002 | -0.002 |
|  | (0.000) | (0.000) | (0.000) | (0.000) | (0.000) | (0.000) |
| Domestic credit | 0.002 | 0.004 | 0.003 | 0.002 | 0.004 | 0.003 |
|  | (0.097) | (0.000) | (0.013) | (0.056) | (0.002) | (0.004) |
| Land area | -0.134 | -0.167 | -0.171 | -0.145 | -0.171 | -0.170 |
|  | (0.003) | (0.000) | (0.000) | (0.002) | (0.000) | (0.000) |
| Trade | -0.001 | -0.001 | -0.002 | -0.001 | -0.001 | -0.001 |
|  | (0.424) | (0.224) | (0.133) | (0.197) | (0.174) | (0.265) |
| Observations | 701 | 701 | 701 | 701 | 701 | 701 |
| Adjusted $R^2$ | 0.595 | 0.582 | 0.587 | 0.591 | 0.584 | 0.588 |
| F-test | 63.747 | 54.714 | 58.351 | 64.782 | 56.226 | 59.548 |
| Log-likelihood | -887.648 | -898.274 | -894.321 | -890.717 | -896.766 | -893.029 |

Note - P-value in parentheses (two-tailed test). Standard errors robust to heteroscedasticity. Year fixed effects, regional fixed effects, and constant included in all models but omitted from table.

**Table 7.** Effects of interaction between disasters and country governance. **3-year lag**
Dependent variable: New business start-up rate

|  | Country Governance | | | | | |
|---|---|---|---|---|---|---|
|  | Voice and Accountability (1) | Political Stability (2) | Government Effectiveness (3) | Regulatory Quality (4) | Rule of Law (5) | Control of Corruption (6) |
| *Explanatory Variables* | | | | | | |
| Natural Disasters $_{t-3}$ | -0.005 | -0.001 | -0.004 | -0.006 | -0.001 | 0.000 |
|  | (0.590) | (0.890) | (0.690) | (0.533) | (0.881) | (0.995) |
| Country Governance $_{t-3}$ | 0.235 | -0.151 | 0.023 | 0.165 | -0.086 | -0.061 |
|  | (0.000) | (0.045) | (0.812) | (0.114) | (0.346) | (0.434) |
| *Interaction Terms* | | | | | | |
| Natural Disasters X Country Governance $_{t-3}$ | 0.004 | 0.013 | 0.021 | 0.020 | 0.016 | 0.027 |
|  | (0.581) | (0.069) | (0.009) | (0.017) | (0.028) | (0.001) |
| *Control Variables* | | | | | | |
| GDP per capita | 0.511 | 0.616 | 0.542 | 0.488 | 0.596 | 0.569 |
|  | (0.000) | (0.000) | (0.000) | (0.000) | (0.000) | (0.000) |
| Cost of start-up procedures | -0.003 | -0.003 | -0.003 | -0.003 | -0.003 | -0.003 |
|  | (0.015) | (0.015) | (0.013) | (0.016) | (0.014) | (0.013) |
| Time required to start a business | -0.004 | -0.003 | -0.002 | -0.003 | -0.003 | -0.003 |
|  | (0.034) | (0.064) | (0.177) | (0.102) | (0.074) | (0.095) |
| Number of start-up procedures | -0.086 | -0.097 | -0.094 | -0.087 | -0.097 | -0.093 |
|  | (0.000) | (0.000) | (0.000) | (0.000) | (0.000) | (0.000) |
| Largest population | -0.076 | -0.042 | -0.062 | -0.028 | 0.009 | -0.075 |
|  | (0.869) | (0.933) | (0.897) | (0.951) | (0.985) | (0.875) |
| GDP growth | 0.023 | 0.020 | 0.021 | 0.022 | 0.021 | 0.020 |
|  | (0.017) | (0.044) | (0.036) | (0.023) | (0.035) | (0.039) |
| Population density | -0.002 | -0.002 | -0.002 | -0.002 | -0.002 | -0.002 |
|  | (0.000) | (0.000) | (0.000) | (0.000) | (0.000) | (0.000) |
| Domestic credit | 0.002 | 0.004 | 0.003 | 0.002 | 0.004 | 0.003 |
|  | (0.088) | (0.000) | (0.021) | (0.071) | (0.002) | (0.008) |
| Land area | -0.134 | -0.167 | -0.169 | -0.145 | -0.167 | -0.167 |
|  | (0.003) | (0.000) | (0.000) | (0.001) | (0.000) | (0.000) |
| Trade | -0.001 | -0.001 | -0.002 | -0.001 | -0.001 | -0.001 |
|  | (0.383) | (0.266) | (0.118) | (0.165) | (0.159) | (0.248) |
| Observations | 701 | 701 | 701 | 701 | 701 | 701 |
| Adjusted $R^2$ | 0.593 | 0.584 | 0.586 | 0.591 | 0.584 | 0.589 |
| F-test | 63.358 | 55.589 | 58.781 | 64.562 | 56.895 | 60.811 |
| Log-likelihood | -888.817 | -897.163 | -894.774 | -891.112 | -897.024 | -892.533 |

Note - P-value in parentheses (two-tailed test). Standard errors robust to heteroscedasticity. Year fixed effects, regional fixed effects, and constant included in all models but omitted from table.

**Table 8.** Effects of interaction between disasters and country governance. **4-year lag**
Dependent variable: New business start-up rate

|  | Country Governance | | | | | |
|---|---|---|---|---|---|---|
|  | Voice and Accountability (1) | Political Stability (2) | Government Effectiveness (3) | Regulatory Quality (4) | Rule of Law (5) | Control of Corruption (6) |
| *Explanatory Variables* | | | | | | |
| Natural Disasters $_{t-4}$ | 0.003 | 0.006 | 0.004 | 0.003 | 0.006 | 0.006 |
|  | (0.750) | (0.487) | (0.650) | (0.752) | (0.533) | (0.479) |
| Country Governance $_{t-4}$ | 0.243 | -0.145 | 0.015 | 0.175 | -0.063 | -0.052 |
|  | (0.000) | (0.062) | (0.873) | (0.077) | (0.462) | (0.486) |
| *Interaction Terms* | | | | | | |
| Natural Disasters X Country Governance $_{t-4}$ | -0.002 | 0.012 | 0.014 | 0.012 | 0.010 | 0.020 |
|  | (0.782) | (0.082) | (0.071) | (0.124) | (0.145) | (0.004) |
| *Control Variables* | | | | | | |
| GDP per capita | 0.530 | 0.626 | 0.566 | 0.507 | 0.606 | 0.581 |
|  | (0.000) | (0.000) | (0.000) | (0.000) | (0.000) | (0.000) |
| Cost of start-up procedures | -0.003 | -0.003 | -0.003 | -0.003 | -0.003 | -0.003 |
|  | (0.018) | (0.016) | (0.015) | (0.019) | (0.016) | (0.015) |
| Time required to start a business | -0.004 | -0.003 | -0.003 | -0.003 | -0.003 | -0.003 |
|  | (0.022) | (0.044) | (0.079) | (0.054) | (0.045) | (0.061) |
| Number of start-up procedures | -0.086 | -0.097 | -0.094 | -0.088 | -0.096 | -0.093 |
|  | (0.000) | (0.000) | (0.000) | (0.000) | (0.000) | (0.000) |
| Largest population | -0.032 | -0.036 | -0.005 | 0.023 | 0.039 | -0.039 |
|  | (0.946) | (0.943) | (0.992) | (0.961) | (0.936) | (0.934) |
| GDP growth | 0.023 | 0.020 | 0.020 | 0.021 | 0.020 | 0.020 |
|  | (0.017) | (0.046) | (0.045) | (0.029) | (0.042) | (0.047) |
| Population density | -0.002 | -0.002 | -0.002 | -0.002 | -0.002 | -0.002 |
|  | (0.000) | (0.000) | (0.000) | (0.000) | (0.000) | (0.000) |
| Domestic credit | 0.002 | 0.004 | 0.003 | 0.002 | 0.004 | 0.003 |
|  | (0.077) | (0.000) | (0.010) | (0.055) | (0.002) | (0.004) |
| Land area | -0.145 | -0.179 | -0.178 | -0.155 | -0.176 | -0.177 |
|  | (0.002) | (0.000) | (0.000) | (0.001) | (0.000) | (0.000) |
| Trade | -0.001 | -0.001 | -0.002 | -0.001 | -0.001 | -0.001 |
|  | (0.382) | (0.279) | (0.144) | (0.169) | (0.171) | (0.232) |
| Observations | 701 | 701 | 701 | 701 | 701 | 701 |
| Adjusted $R^2$ | 0.593 | 0.584 | 0.586 | 0.591 | 0.584 | 0.589 |
| F-test | 63.358 | 55.589 | 58.781 | 64.562 | 56.895 | 60.811 |
| Log-likelihood | -888.817 | -897.163 | -894.774 | -891.112 | -897.024 | -892.533 |

Note - P-value in parentheses (two-tailed test). Standard errors robust to heteroscedasticity. Year fixed effects, regional fixed effects, and constant included in all models but omitted from table.